 	\documentclass[aps,prb,twocolumn,	floatfix,showpacs]{revtex4}
 \pdfoutput=1
\usepackage{graphicx} 
\usepackage{calc} 
\usepackage{amsfonts}  
\usepackage{amsmath} 

\begin{document}
\newcommand{\kw}{ \bar{k} }
\newcommand{\qq}{  \bar{q}_0  }
\newcommand{\qw}{ \bar{q} }
\newcommand{\G}{ \bar{G} }
\def\sumslashD{\mathop{\sum \kern-1.4em -\kern 0.5em}}
\def\sumslash{\mathop{\sum \kern-1.2em -\kern 0.5em}}
\def\intslash{\mathop{\int \kern-0.9em -\kern 0.5em}}
\def\intslashD{\mathop{\int \kern-1.1em -\kern 0.5em}}

\newcommand{\hc}{{\mathrm{H.c.}}}	
\newcommand{\potloc}{\mathcal U}	
\newcommand{\hamil}{{H}}		
\newcommand{\paren}[1]{\left(#1\right)}	

\title{Role of electron-phonon interaction in a magnetically driven mechanism for superconductivity}        
 
\author{H. Bakrim  and C. Bourbonnais}
\affiliation{ Regroupement Qu\'ebecois sur les Mat\'eriaux de Pointe,
  D\'epartement de physique, Universit\'e de Sherbrooke, Sherbrooke,
  Qu\'ebec, Canada, J1K-2R1 }

\date{\today}

\begin{abstract} 
We use  the renormalization group method to examine  the effect  of    phonon mediated  interaction  on {\it d}-wave superconductivity, as driven  by spin fluctuations  in  a quasi-one-dimensional electron system.    The influence of a   tight-binding electron-phonon interaction on the spin-density-wave and {\it d}-wave superconducting instability lines  is calculated  for arbitrary temperature, phonon frequency and antinesting of the Fermi surface.
   The  domain of electron-phonon coupling  strength where  spin-density-wave order  becomes unstable against  the formation of a  bond-order-wave or  Peierls state is determined at weak antinesting. We show the existence of a  positive isotope effect   for  spin-density-wave and {\it d}-wave superconducting critical temperatures  which scales with the antinesting distance from quantum critical point where the two instabilities merge.  We single out a low phonon frequency zone where the bond-oder-wave ordering gives rise to  triplet {\it f}-wave superconductivity under nesting alteration,  with both orderings displaying  a negative isotope effect. We also study the   electron-phonon strengthening   of spin fluctuations at the origin of extended quantum criticality in the metallic phase above superconductivity.  The impact of our results on   quasi-one-dimensional organic conductors like the Bechgaard salts where a Peierls distortion is absent and superconductivity emerges near a spin-density-wave state under pressure is emphasized.
    \end{abstract}
\maketitle 

\section{Introduction}
  The role of electron-phonon coupling in strongly correlated electron systems is an issue  of growing importance  in   materials  where unconventional superconductivity is found to compete with various forms of electronic states.\cite{Gurnnasson08,Capone10,Lee14} A point at issue in the quest of understanding the origin of superconductivity  is the extend to which the electron-phonon interaction can  influence  and even modify the nature of Cooper pairing  when electrons   are strongly correlated, especially through magnetism.  In this work we shall  focus on this issue  in  quasi-one-dimensional molecular superconductors where superconductivity takes place in the close proximity of anti ferromagnetism, as best exemplified in  the  Bechgaard salts series of organic superconductors.\cite{Bourbon08}

Since the discovery of  superconductivity (SC) in the Bechgaard salts   [(TMTSF)$_2X$] series\cite{Jerome80}, much of the attention paid to the mechanism  of Cooper pairing  has mostly focused  on  models of electrons  with purely repulsive interactions\cite{Bourbon08,Emery82,Emery86,Caron86B,BealMonod86,Scalapino86,Shimahara89,Mazumdar00,Kuroki01,Fuseya05b,Fuseya05c,Friedel06}. On empirical grounds, this  has been amply supported  by the ubiquity  of spin-density-wave (SDW) correlations   nearby the superconducting state  when   pressure \cite{Jerome82,Azevedo84,Vuletic02,DoironLeyraud09},  temperature  \cite{Bourbon84,Creuzet87b}, or  even  magnetic field is varied \cite{Wang93,Cooper89}.
As one moves  along the temperature axis, for example, and enters the metallic state, important SDW fluctuations are   found to govern  properties of  the normal phase, giving rise, for instance, to a huge  enhancement of the NMR spin relaxation rate and to linear-$T$ resistivity term  over a wide temperature  interval above the critical temperature $T_c$ for superconductivity  \cite{Creuzet87b,Wu05,DoironLeyraud09}. 

 Besides the nesting of the Fermi surface, repulsive interactions  are an essential  component of  SDW correlations \cite{Yamaji82,Emery82,Gutfreund8183,Hasegawa86b}. They have  become inescapable  ingredients  of  the model description of superconductivity in these materials. 
In this regard, the  quasi-one dimensional electron gas model  with the aid of the renormalization group (RG) method have played an important part in the description of these low-dimensional electron systems. In the repulsive sector, it proved particularly generic of the SDW-to-{\it d}-wave SC (SC-{\it d}) sequence of instabilities  when the amplitude of the next-to-nearest neighbor interchain hopping, $t_\perp'$,   called the anti nesting parameter, is tuned to    simulate   pressure effects  on spin fluctuations  responsible for superconducting pairing interaction \cite{Duprat01,Nickel0506}.  The   approach has also shown how the   constructive interference  between spin fluctuations and Cooper pairing  can explain the existence of a   Curie-Weiss temperature dependence  of the SDW   correlation length,  which is a key factor in the enhancement of  the NMR  relaxation rate  and the  linear-$T$ component in resistivity over the whole pressure interval where superconductivity is present   \cite{DoironLeyraud09,Bourbon09,Sedeki12,Meier13}.

However, in view of the complex molecular structure of systems like the Bechgaard salts,  the repulsive electron  gas model must    be regarded as an idealization. 
   It ignores primarily   the interaction of electrons with  low-energy phonon modes of the lattice.     Early  x-ray diffuse scattering experiments in (TMTSF)$_2$PF$_6$ and (TMTSF)$_2$ClO$_4$ compounds  did reveal  the existence of  such a coupling, under the guise of lattice fluctuations  at the one-dimensional (1D)  wave vector $2k_F$ of the electron gas  ($k_F$ being the longitudinal Fermi wave vector) \cite{Pouget82,Pouget12}.  The lattice fluctuations    remain   regular  in temperature for the Bechgaard salts, in contrast to so many molecular chain systems where it terminates in  a   Peierls  [ bond-order-wave (BOW) ] distorted state. Although the reason  for this remains large ly unexplained \cite{Emery83,Pouget12}, the presence of  2$k_F$ lattice fluctuations is  direct evidence of a finite  (momentum dependent) coupling  between electrons and phonons,    a  consequence   of  the modulation of  tight-binding electron band parameters  by lattice vibrations.

This points  at  the    impact a retarded  [phonon-mediated (Ph-M)] interaction of that kind can have  on the properties of the electron gas when the mechanism for  Cooper  pairing is  magnetically driven. Whether it   is    detrimental to SDW and SC-{\it d} correlations, as one would naturally  expect  if the electron-phonon  interaction was  taken in isolation 
 \cite{Bakrim10},   or   on the contrary, if it   becomes    a  factor of reinforcement  when  it is  subordinate to repulsive interactions is an open question.     
The  latter possibility  can provide new insight as to the conditions prevailing  in weakly dimerized systems like the Bechgaard salts that   make  SDW predominate over the  Peierls phenomena. 

  More generically, it can clarify   how a momentum dependent electron-phonon  interaction can be actively involved in the occurrence of superconductivity near magnetism.  It  can also shed light on the possibility of   a     
    positive isotope effect   for the temperature scale of  instabilities against SDW and SC-{\it d} orderings as a function of phonon frequency.  
    Reinforcement could also extend relatively far in the   metallic phase  by enhancing spin fluctuations as quantum critical effects  due to interfering SC-{\it d}  and SDW instabilities \cite{Sedeki12}. 
    
    These possibilities are important in the context of  other unconventional superconductors, in particular high-$T_c$  cuprates \cite{Lanzara01,Gweon04,Iwasawa08,Crawford90},  where  they framed  a significant part of the debate surrounding the relative importance  of Coulomb and electron-phonon interactions when superconductivity takes place in the proximity of antiferromagnetism  \cite{Bauer10,Klironomos06,Tsai05,Sangiovanni05,Gurnnasson08,Honerkamp07,Huang03,Sandvik04b,Andergassen01,Bulut96} and charge-density-wave ordering \cite{Chang12,LeTacon14}. Its transposition  in  quasi-one-dimensional superconductors like the Bechgaard salts close to a SDW instability has  remained  essentially unexplored since the very first attempts to reconcile electron-electron and electron-phonon interactions in the framework of mean-field  theory of competing  magnetism and superconductivity  \cite{Gutfreund8183}. 

 In this work we shall  address this  problem in the weak-coupling framework of the RG approach to  the quasi-1D electron gas model. The model  is  extended to include  both direct and  momentum-dependent Ph-M  electron-electron  interactions in the the study of interfering (electron-electron) Cooper and (electron-hole) density-wave pairings at arbitrary phonon frequency $\omega_D$.   The RG calculations will be carried out  at finite temperature $T$, which brings additional difficulties in the presence of retarded interactions. This turns out to be  required  when    antinesting is present. Actually,   a finite $t_\perp'$  breaks  the usual  correspondence between $T$ and the scaled cut-off energy $\Lambda(\ell)$   from the Fermi surface that generates the RG flow.  The flow will be  then  conducted at arbitrary temperature for interactions with     momentum dependence   along the Fermi surface  and a finite set of Matsubara frequencies. This finite$-T$  RG  procedure with momentum and frequency variables has been worked out   recently for systems where Ph-M  interactions are predominant, a situation relevant to    competing charge-density-wave and {\it s}-wave SC instabilities  away from half-filling \cite{Bakrim10}.    It is extended here to weakly dimerized chains systems like the Bechgaard salts where repulsive  interactions are dominant  and half-filling Umklapp scattering is finite \cite{Emery82,Sedeki12,Bourbon09}.         

  The results put forward below  show that the modulation of tight-binding electron band by acoustic lattice vibrations introduces effective   Ph-M interactions with a very characteristic dependence  on longitudinal electron momentum and momentum transfer of scattered electrons.  The dependence affects the RG flow and produces  a     low-energy downward screening of the repulsive backward scattering  and  an enhancement of both repulsive forward and   Umklapp scattering   terms. These  effects  are $\omega_D$ dependent,  concurring to boost antiferromagnetic exchange between itinerant electrons which primarily  reinforces the    SDW  instability line and in turn the magnetically driven  SC-{\it d} in  the phase diagram. The impact of retardation  generates  a positive isotope effect  whose  amplitude peaks at the critical   strength of antinesting where SDW    and SC-{\it d} instabilities lines meet and  their  constructive  interference is the  strongest. Above a definite strength of electron-phonon interaction, the SDW becomes unstable against the formation of a BOW distorted state  and   triplet {\it f}-wave superconductivity   if    antinesting and retardation are sufficiently high. The latter states are both characterized by a negative isotope effect, as a result  of antiadiabaticity. 
  
  The boost of  exchange by electron-phonon interaction  is not limited to the transition lines, but is  manifest   in the metallic phase  where it feeds deviations to Fermi liquid behavior    at the origin  of extended quantum criticality  in the normal phase\cite{Sedeki12,Bourbon09,DoironLeyraud09}. The latter can be followed through the reinforcement of the Curie-Weiss  behavior  of the SDW  susceptibility which is   correlated to $\omega_D$ and   antinesting $t_\perp'$  in the whole range   where superconductivity is present.
  
  In Sec. II we introduce the quasi-1D electron gas model which  is extended to include the tight-binding electron-phonon interaction term. In Sec.~III the one-loop RG flow equations for the different electron-electron  vertices and relevant response functions  are given and integrated into the determination of the phase diagram at arbitrary anti nesting and phonon frequency. Their integration is carried out in Sec.~IV and leads to  the determination of the phase diagrams, isotope effects, and spin  fluctuations in the normal state. In Sec.~V, we discuss the implications of our results in the description of unconventional superconductors like the Bechgaard salts and conclude this work.\section{The model }
For  a linear array of $N_\perp$ chains of length $L$,   the Hamiltonian of the quasi-1D electron gas   with electron-phonon coupling  is given by    
\begin{align} 
\label{Hamiltonian}
&  H   =    H_{\rm p}^0 + H_{\rm ep} +      \sum_{p,\mathbf{k},\sigma}   E_p(\mathbf{k})\, c^\dagger_{p,\mathbf{k},\sigma}c_{p,\mathbf{k},\sigma} \cr
  & +   {\pi v_F\over LN_\perp} \sum_{\{\mathbf{k},\sigma\}} \,   \big[\, g_1\, c^\dagger_{+,\mathbf{k}_4,\sigma_1}  c^\dagger_{-,\mathbf{k}_3,\sigma_2}   c_{+,\mathbf{k}_2,\sigma_2}c_{-,\mathbf{k}_1,\sigma_1}  \cr    
& \hskip 2truecm    +   \ g_2 \, c^\dagger_{+,\mathbf{k}_4,\sigma_1}c^\dagger_{-,\mathbf{k}_3,\sigma_2}c_{-,\mathbf{k}_2,\sigma_2}c_{+,\mathbf{k}_1,\sigma_1}   \cr
  &\hskip 2truecm        + {1\over 2}g_3 \, \big( c^\dagger_{+,\mathbf{k}_4,\sigma_1}c^\dagger_{+,\mathbf{k}_3,\sigma_2}c_{-,\mathbf{k}_2,\sigma_2}c_{-,\mathbf{k}_1,\sigma_1}  \cr
  & \hskip 2truecm + \mathrm{H.c} \big)\big]\delta_{\mathbf{k}_1+\mathbf{k}_2= \mathbf{k}_3+\mathbf{k}_4 (\pm\mathbf{G}) },
 \end{align}
In the purely electronic  part that has  been made explicit, the operator $c^\dagger_{p,\mathbf{k},\sigma}$ $(c_{p,\mathbf{k},\sigma}$) creates (destroys) a right ($p=+$) and left $(p=-)$ moving electron of  wave vector  $\mathbf{k}=(k,k_\perp)$  and spin $\sigma$.   The free part is  modeled  by the anisotropic  one-electron energy spectrum in two dimensions, 
\begin{equation}
\label{spectrum}
\begin{split}
E_p(\mathbf{k} ) =  v_F(pk-k_F)  + \epsilon(k_\perp),
\end{split}
\end{equation}
where
\begin{equation}
\label{ }
\epsilon(k_\perp) = - 2t_\perp\cos k_\perp -2t_\perp'\cos 2k_\perp.
\end{equation}
The  longitudinal part has been linearized around the longitudinal Fermi  wave vector given by $pk_F= \pm \pi/2$ for a dimerized chain with one electron per dimer. The longitudinal Fermi velocity is $v_F= 2t $ where $t$ is the average nearest-neighbour hopping. Here     $t_{\perp}$  is  the  nearest-neighbor hopping integral  in the   perpendicular direction and $t_{\perp }'$ is a second nearest-neighbor hopping   paramaterizing   deviations to perfect nesting at $\mathbf{q}_0=(2k_F,\pi)$, which  simulates the most important effect of pressure in our model.  The   quasi-1D anisotropy  of the spectrum is  \hbox{$E_F\simeq 15 t_{\perp }$}, where $E_F = v_Fk_F\simeq 3000 $K is the longitudinal Fermi energy congruent with the range  found in the Bechgaard salts\cite{Grant82,Ducasse86,LePevelen01}; $E_F$ is taken as half the bandwidth  cutoff $E_0=2E_F$ in the model.    In the framework of the electron gas model\cite{Solyom79,Emery79}, the interacting part of the Hamiltonian  is described by the bare backward, \hbox{$g_1\equiv g_1(+k_F,-k_F ;+k_F,-k_F)$},  and forward,  $g_2\equiv g_2(+k_F,-k_F ;-k_F,+k_F)$, scattering amplitudes between right and left moving electrons defined  on the  1D Fermi surface.  The half-filling character of the band -- a consequence of a small dimerization of the chains -- gives rise to  Umklapp scattering of bare amplitude $g_3\equiv g_3( \pm k_F, \pm k_F; \mp k_F, \mp    k_F)$,  and for which momentum conservation involves the longitudinal reciprocal lattice vector $\mathbf{G}=(4k_F,0)$.  Within the electron gas model,  the deviation $k\pm k_F$ of longitudinal momentum  with respect to the Fermi points in the scattering amplitudes are irrelevant in the RG sense and can be  neglected\cite{Solyom79,Emery79,Menard11}.  All   couplings are  normalized by $\pi v_F$ and are initially independent of transverse momenta $k_{\perp i}$, but acquire such an dependence along the RG flow. This momentum dependence refers to the angular dependence along the Fermi surface.

Regarding the values taken by the  interaction parameters throughout  the present calculations, we shall take \hbox{ $g_1=g_2/2 \simeq 0.32$}  and $g_3\simeq 0.025$, which follows from the phenomenological analysis of previous works that fixes  their amplitude  from different experiments in the weakly dimerized systems like  the Bechgaard salts\cite{Bourbon09,Sedeki12}. This pertains to a range of couplings generic of the interplay between SDW and SCd orders as a function of antinesting.

The electron-phonon part of the hamiltonian (\ref{Hamiltonian}) follows from the modulation of the longitudinal hopping integral by acoustic phonons in the tight-binding approximation \cite{Su80B}. It  reads
\begin{align}
\label{ }
& \ \ \ \ \ \ \ H_{\rm p}^0  +  H_{\rm ep}=   \sum_{\mathbf{q},\nu} \omega_{\mathbf{q},\nu}\Big(b^\dagger_{{\mathbf{q},\nu}}b_{\mathbf{q},\nu} + {1\over2}\Big)   \cr   
&\  +  (LN_\perp)^{-{1\over 2}}\sum_{p,\sigma,\nu}\sum_{\mathbf{k},\mathbf{q}} g_\nu( {k}, {q})c^\dagger_{p,\mathbf{k} + \mathbf{q},\sigma}c_{-p,\mathbf{k},\sigma}(b^\dagger_{\mathbf{q},\nu} + b_{ {-\mathbf{q},\nu}})\cr
\end{align}
where $\nu$ is related to the different polarization of   acoustic phonons. For phonons of interest propagating parallel to the chains axis, we have 
\begin{equation}
\omega_{q,\nu}   =   \omega_\nu\big|\sin{q\over 2}\big|, 
\end{equation}
for the phonon spectrum and
\begin{equation}
g_\nu(k,q)   =   i 4{\lambda_\nu\over \sqrt{2M \omega_\nu}} \sin{q\over 2} \cos\big(k + {q\over 2}\big),
\label{SSH}
\end{equation} 
  for the electron-phonon matrix element, which depends on both electron momentum $k$ and momentum transfer $q$.  The coupling amplitude \hbox{$\lambda_\nu= \nabla t\cdot {\bf e}_\nu$} is expressed in terms of the spatial variation of longitudinal hooping integral and the unit vector ${\bf e}_\nu$ of the lattice displacement;   $\omega_\nu = 2 \sqrt{\kappa_\nu/M}$ is the Debye frequency for the acoustic branch $\nu$, and $M$ is the mass of molecular unit. The bandwidth of  acoustic branches in the molecular systems like the Bechgaard salts does not exceed $\omega_\nu \sim100$~K~\cite{Krauzman86,Homes89,Pouget76}. We shall consider  in the following the interval normalized phonon frequency  $0< \omega_D/t_\perp\le 0.5$. 

For the partition function $Z$, it is  straightforward to proceed to the  partial trace of   harmonic  phonon degrees of freedom and express the partition function,
$$
Z= \int\!\!\int {\mathfrak D}\psi^* {\mathfrak D}\psi^* e^{S_0 + S_I},
$$ 
   as a functional integral  over the fermion anti commuting fields $\psi^{(*)}$. The bare action in the Matsubara-Fourier space is given by
\begin{equation}
\label{ }
S_0[\psi^*,\psi] =  \sum_{\bar{\bf{k}},p,\sigma} [G^0_p(\bar{\mathbf{k}})]^{-1} \psi^*_{p,\sigma}(\bar{\mathbf{k}})\psi_{p,\sigma}(\bar{\mathbf{k}})
\end{equation} 
where $\bar{\mathbf{k}}=(\mathbf{k},\omega_n=\pm \pi T, \pm 3\pi T, \ldots)$ and 
\begin{equation}
\label{ }
G^0_p(\bar{\mathbf{k}}) = [i\omega_n- E_p(\mathbf{k})]^{-1}
\end{equation}
is the bare fermion propagator. The interacting part of the action is of  the form 
\begin{widetext}
\begin{align}
\label{SI}
S_I[\psi^*,\psi]= - {T\over L N_\perp}\pi v_F \sum_{\{\bar{\mathbf{k}},\sigma\}}   \{& \ g_{1}(\bar{k}_1,\bar{k}_2,\bar{k}_3,\bar{k}_4) \psi^*_{+,\sigma_4}(\bar{\mathbf{k}}_4)\psi^*_{-,\sigma_3}(\bar{\mathbf{k}}_3)\psi_{+,\sigma_2}(\bar{\mathbf{k}}_2)\psi_{-,\sigma_1}(\bar{\mathbf{k}}_1)  \cr
+ & \ g_{2}(\bar{k}_1,\bar{k}_2,\bar{k}_3,\bar{k}_4) \psi^*_{+,\sigma_4}(\bar{\mathbf{k}}_4)\psi^*_{-,\sigma_3}(\bar{\mathbf{k}}_3)\psi_{-,\sigma_2}(\bar{\mathbf{k}}_2)\psi_{+,\sigma_1}(\bar{\mathbf{k}}_1) \cr
+ & {1\over 2} [g_{3}(\bar{k}_1,\bar{k}_2,\bar{k}_3,\bar{k}_4) \psi^*_{+,\sigma_4}(\bar{\mathbf{k}}_4)\psi^*_{+,\sigma_3}(\bar{\mathbf{k}}_3)\psi_{-,\sigma_2}(\bar{\mathbf{k}}_2)\psi_{-,\sigma_1}(\bar{\mathbf{k}}_1) + {\rm c.c.}] \}  \delta_{\bar{\mathbf{k}}_1+ \bar{\mathbf{k}}_2, \bar{\mathbf{k}}_3+ \bar{\mathbf{k}}_4   (\pm \bar{\mathbf{G}}) }
\end{align}
\end{widetext}
where $\bar{k}_i\equiv (k_{\perp i},\omega_{n i})$ and $\overline{\mathbf{G}}= (4k_F,0,0)$ for Umklapp scattering. The  amplitude of the bare effective backscattering is given by
\begin{align}
 g_{1}(\bar{k}_1,\bar{k}_2,\bar{k}_3,\bar{k}_4) =  & \  g_1 - \sum_\nu  {2\over \pi v_F\omega_\nu}\cr &  \ \ \ \times { g_\nu(k_F,-2k_F) g_\nu(-k_F,2k_F)\over
1+  (\omega_{n 3}-\omega_{n 1})^2/\omega_\nu^2} \cr
 \equiv  & \ g_1 +  {g_{\rm ph}\over 1 + (\omega_{n 3}-\omega_{n 1})^2/\omega_D^2}, 
 \label{g1}
  \end{align}
where   in the   electron gas model scheme   the interactions are defined  on the 1D Fermi points with the electron-phonon matrix element  evaluated   at $k=\pm k_F$ and momentum transfer $ q=\pm 2k_F$. Here we have defined the Debye frequency $\omega_D = \langle \omega_{2k_F,\nu}\rangle$, as the average phonon frequency over  the different   branches at the zone edge. We can define  an attractive  contribution  from all  acoustic branches of  normalized amplitude 
\begin{equation}
\label{gph}
g_{\rm ph} = -4\sum_\nu \lambda^2_\nu /(\pi v_F \kappa_\nu) .
\end{equation}  
  Likewise, for the amplitude of the effective forward scattering, we have    
  \begin{align}
  g_{2}(\bar{k}_1,\bar{k}_2,\bar{k}_3,\bar{k}_4) =  &\,  g_2 - {2\over \pi v_F} \sum_\nu \omega_{0,\nu}{g_\nu(k_F,0) g_\nu(-k_F,0)\over
\omega_{0,\nu}^2 + (\omega_{n3}-\omega_{n1})^2 } \cr 
 =  & \ g_2,
 \label{g2}
\end{align}
which remains unaffected by   phonons at vanishing momentum transfer. Finally, for the bare   Umklapp term in the presence of phonons, we have
\begin{align}
 g_{3}(\bar{k}_1,\bar{k}_2,\bar{k}_3,\bar{k}_4) =  &\,  g_3 - \sum_\nu  {2\over \pi v_F \omega_\nu} \cr & \ \ \ \times { g_\nu(k_F,2k_F) g_\nu(k_F,-2k_F)\over
1+  (\omega_{n 3}-\omega_{n 1})^2/\omega_\nu^2} \cr
 \equiv  & \ g_3 +  {\eta |g_{\rm ph}|\over 1 + (\omega_{n 3}-\omega_{n 1})^2/\omega_D^2},
 \label{g3}
  \end{align}
  which, in contrast to normal backscattering, gives rise to a retarded repulsive contribution,   as a result of the $k$ dependence of the electron-phonon tight binding matrix element (\ref{SSH}).   Here $\eta$ is a reduction factor that takes into account  the weak dimerization of the chains. For simplicity we shall take $\eta=g_3/g_1 (= \Delta_D/E_F\ll 1) $ (see also Ref.~\cite{Gutfreund8183}). 
  
The dependence  of the above bare retarded couplings on  both  longitudinal  $ k$  and   momentum transfer $q$  will play an important role in their RG  flow at low energy.
\section{Renormalization group equations}
We use  the finite temperature momentum-frequency RG scheme introduced in Ref.\cite{Bakrim10}. In the partition function we proceed    to the successive integration  of electron states in the energy shell $\Lambda(\ell) d\ell$ at energy distance  $ \pm \Lambda(\ell) =\pm E_Fe^{-\ell}$  from the  Fermi surface,     where $\ell\in[0,\infty)$. For the $k_\perp$-momentum dependence of the scattering amplitudes  on each  Fermi sheet, a constant energy surface in the Brillouin zone is separated into 12 patches, inside    which   the couplings are considered constant in the loop integration \cite{Nickel0506}. The number of patches   is sufficient to take into account the non perturbative effect of warping of the Fermi surface and the antinesting term $t_\perp'$.  Regarding the frequency dependence  we have considered a finite number of $N_\omega= 14$  Matsubara frequencies $\omega_{n} $ ($ -7\le n\le 6)$, within a mean-field single patch scheme  for  the loop frequency variable   as described below. 
The flow equations read 
\begin{align}
\label{RGg1}
& \partial_\ell g_{1}(\bar{k}_1,\bar{k}_2,\bar{k}_3,\bar{k}_4) =  {1\over 2\pi} \int dk_\perp I_P(k_\perp, \bar{q}_P) \cr
\times  \Big[ &\, \epsilon_P\langle g_1 (\bar{k}_1,\bar{k},\bar{k}_P,\bar{k}_4) g_1 (\bar{k}_P,\bar{k}_2,\bar{k}_3,\bar{k}) \rangle  \cr  
&+ \epsilon_{P,v} \langle g_2 (\bar{k}_1,\bar{k},\bar{k}_{4},\bar{k}_P) g_1 (\bar{k}_P,\bar{k}_2,\bar{k}_3,\bar{k}) \rangle \cr
&+ \epsilon_{P,v} \langle g_1 (\bar{k}_1,\bar{k},\bar{k}_P,\bar{k}_{4}) g_2 (\bar{k}_P,\bar{k}_2,\bar{k},\bar{k}_3) \rangle\Big]_{}    \cr
&+ \epsilon_P     \langle g_3 (\bar{k}_1,\bar{k},\bar{k}_3,\bar{k}_P') g_3(\bar{k}_P',\bar{k}_2,\bar{k},\bar{k}_4) \rangle \cr
&+ \epsilon_{P,v} \langle g_3 (\bar{k},\bar{k}_1,\bar{k}_3,\bar{k}_P') g_3(\bar{k}_P',\bar{k}_2,\bar{k},\bar{k}_4) \rangle \cr
&+ \epsilon_{P,v} \langle g_3 (\bar{k}_1,\bar{k},\bar{k}_P',\bar{k}_3) g_3(\bar{k}_P',\bar{k}_2,\bar{k},\bar{k}_4) \rangle \Big]_{}  \cr 
&+ {1\over 2\pi} \int dk_\perp I_C(k_\perp, \bar{q}_C) \cr
\times \Big[ &  \epsilon_{C} \langle g_1 (\bar{k}_1,\bar{k}_2,\bar{k},\bar{k}_C) g_2 (\bar{k},\bar{k}_C,\bar{k}_4,\bar{k}_3) \rangle  \cr
    &+   \epsilon_{C}  \langle g_2 (\bar{k}_1,\bar{k}_2,\bar{k}_C,\bar{k}) g_1 (\bar{k},\bar{k}_C,\bar{k}_3,\bar{k}_4) \rangle\Big],
    \end{align}
    \begin{align}   
 &  \partial_\ell g_{2}(\bar{k}_1,\bar{k}_2,\bar{k}_3,\bar{k}_4) =   {1\over 2\pi} \int dk_\perp  I_P(k_\perp, \bar{q}_P') \cr
\times  & \Big[\  \epsilon_{P,l} \langle g_2 (\bar{k}_1,\bar{k},\bar{k}_3,\bar{k}'_P) g_2 (\bar{k}'_P,\bar{k}_2,\bar{k},\bar{k}_4) \rangle  \cr 
& + \epsilon_{P,l} \langle g_3(\bar{k}_1,\bar{k},\bar{k}_P,\bar{k}_4) g_3(\bar{k}_P,\bar{k}_2,\bar{k}_3,\bar{k}) \rangle\Big]_{}\cr  
& +  {1\over 2\pi} \int dk_\perp  I_C(k_\perp, \bar{q}_C) \cr
 \times \Big[& \,\epsilon_{C}  \langle g_1 (\bar{k}_1,\bar{k}_2,\bar{k},\bar{k}_C) g_1 (\bar{k},\bar{k}_C,\bar{k}_4,\bar{k}_3) \rangle \cr
       &\quad  + \epsilon_{C} \langle g_2 (\bar{k}_1,\bar{k}_2,\bar{k}_C,\bar{k}) g_2 (\bar{k},\bar{k}_C,\bar{k}_3,\bar{k}_4) \rangle\Big],
       \label{RGg2}
       \end{align}
and 
\begin{align}
& \partial_\ell g_3(\bar{k}_1,\bar{k}_2,\bar{k}_3,\bar{k}_4) = {1\over 2\pi} \int dk_\perp I_P(k_\perp, \bar{q}_P) \cr
\times  & 2 \Big[ \ \epsilon_P\langle g_1 (\bar{k}_1,\bar{k},\bar{k}_3,\bar{k}_P') g_3 (\bar{k}_P',\bar{k}_2,\bar{k},\bar{k}_4) \rangle  \cr  
&+ \epsilon_{P,v} \langle g_1 (\bar{k}_1,\bar{k},\bar{k}_3,\bar{k}_P') g_3 (\bar{k}_P',\bar{k}_2,\bar{k}_4,\bar{k}) \rangle \cr
&+ \epsilon_{P,v} \langle g_2 (\bar{k},\bar{k}_1,\bar{k}_3,\bar{k}_P') g_3 (\bar{k}_P',\bar{k}_2,\bar{k},\bar{k}_4) \rangle\Big] \cr
&+ {1\over 2\pi} \int dk_\perp I_P(k_\perp, \bar{q}_P')\cr \times & \ \ \ 2 \epsilon_{P,l} \langle g_2 ((\bar{k},\bar{k}_1,\bar{k}_3,\bar{k}_P') g_3(\bar{k}_P',\bar{k}_4,\bar{k}_2,\bar{k})    \rangle
\label{RGg3}
\end{align}
 These consist of  closed loop ($\epsilon_P=-2$), vertex corrections  ($\epsilon_{P,v}=1$) and ladder ($\epsilon_{P,l}=1$) diagrams of the ${\bf q}_P$ electron-hole (Peierls) pairing,   which combine with  the  ladder diagrams ($\epsilon_C=-1$) of the electron-electron (Cooper) pairing. Here  $\bar{k}_P= \bar{k} + \bar{q}_P$   ,$\bar{k}'_P= \bar{k} + \bar{q}_P'$ and  $\bar{k}_C= -\bar{k} +  \bar{q}_C$, where  $ \bar{q}_{P,C}= (q_{\perp P,C}, \omega_{P,C})$   corresponds to the Peierls $ \bar{q}_P=\bar{k}_1 -\bar{k}_4$,  $ \bar{q}'_P=\bar{k}_1 -\bar{k}_3$ and Cooper $\bar{q}_C=\bar{k}_2 +\bar{k}_1$ variables. 
 
  In the above equations, each diagram singles out a discrete frequency convolution of the form ${\cal D}_{P,C} =\sum_{\omega_n} g_i \circ g_j \circ {\cal L}_{P,C}$, between the coupling products and the electronic Peierls (Cooper)  loop derivative ${\cal L}_{P,C}= T\partial_\ell G^0_+(\bar{k}+ \bar{q}_{P,C})G^0_-(\pm\bar{k})$. The exact infinite frequency summation at  arbitrary  $T$ is computationally out of reach.  It will be approximated, however, according  to the following decoupling scheme in which the  frequency summation in electronic loops  is decoupled  from  the interactions, the latter having a frequency dependence mainly concentrated  below $\omega_D$, as shown by  the phonon propagators of Eqs.~(\ref{g1})-(\ref{g3}).

We therefore  use  a  mean-field  scheme in which  ${\cal D}_{P,C} \to \langle g_i\circ g_j\rangle   \sum_{\omega_n}{\cal L}_{P,C}$, where $\langle \cdots \rangle  ={N_\omega}^{-1}\sum_{n} \cdots$, stands as an average  of the coupling part   over a finite set of $\omega_n$. As for the electronic part, corresponding to the $\ell$ derivative of the     Cooper and Peierls loops,  $I_{P,C}= \sum^{+\infty}_{n=-\infty}{\cal L}_{P,C}$, the  summation over all frequencies can be computed exactly giving an explicit    dependence on  temperature $T$ and external electronic frequencies $\omega_{P,C}$. This yields 
 \begin{align}
\label{IPC}
   I_{P,C}&(k_\perp,\bar{q}_{P,C})=  \sum_{\nu=\pm 1} \theta[|E_0(\ell)/2 + \nu A_{P,C}|- E_0(\ell)/2]   \cr
    &\times {1\over 4} \left[\tanh {E_0(\ell) + 2 \nu A_{P,C}\over4T} + \tanh {E_0(\ell)\over 4T}\right]\cr
    &\ \ \  \ \ \times \frac{(E_0(\ell) + \nu A_{P,C})E_0(\ell)}{(E_0(\ell) + \nu A_{P,C})^{2}+ \omega^{2}_{P,C}},  
\end{align} where $\omega_P=\omega_{n3}-\omega_{n1}$, and
\begin{equation}
A_P= -\varepsilon(k_\perp) -\varepsilon(k_\perp+q_{\perp P}), 
\end{equation}
for the Peierls channel;  $\omega_C=\omega_{n1}+ \omega_{n2}$, and
\begin{equation}
A_C=   -\varepsilon(k_\perp) + \varepsilon(k_\perp+ q_{\perp C}),  
\end{equation}
for the   Cooper  channel. Here $\theta[x]$ is the step function ($\theta[0]\equiv {1\over2}$).   At finite temperature, the above decoupling scheme with the number of frequencies  $(=14)$ and momentum patches $(=12)$ retained  corresponds to the solution of $(14 \times 12)^3$ $\sim   1.2\times 10 ^6$ coupled RG flow equations governed by Eqs.~(\ref{RGg1}-\ref{RGg3}), a number that has been reduced by various symmetries of the coupling constants with respect to frequencies and transverse momenta. 

  The approximation  can reasonably well take into account retardation effects for a  ratio $\omega_D/\pi T$ that is not too large. It represents  a good compromise 
between exacting computing time and reproducing the
results known for either the non-retarded case in quasi-1D
\cite{Nickel0506,Bourbon09,Sedeki12} or the  quantum corrections to the BOW ordering in  a pure electron-phonon problem in one
dimension \cite{Bakrim07,Bakrim10}.

The  nature of instabilities of the electron gas and their critical temperatures, $T_\mu$, are best studied from the   susceptibilities $\chi_\mu$. For the coupled electron-phonon model under consideration, only  superconducting and staggered density-wave susceptibilities   present a singularity as a function of antinesting and electron-phonon interaction strength. In the static limit, these are defined  by  
\begin{equation}
\label{ }
\pi v_F\chi_\mu(\bar{q}_\mu^0) = {1\over 2\pi} \!\! \int \!\! dk_\perp\!\!\int_\ell \langle z^2_\mu (\bar{k} + \bar{q}_\mu^0)\rangle I_{P,C}(k_\perp +  \bar{q}_\mu^0)d\ell,
\end{equation}
 where the vertex parts $z_\mu$ are governed by    one-loop flow equations. In the density-wave   channel, we shall consider
  \begin{align}
\label{RGSDW}
& \partial_\ell  z_{\rm SDW}(\bar{k}+ \bar{q}_P^0) =   {1\over 2\pi} \int dk'_\perp I_P(k_\perp',\bar{q}^0_P) z_{\rm SDW}(\bar{k}' + \bar{q}_P^0)\cr 
&\! \! \times \! \big\langle  [  \epsilon_{P,l} g_3(\bar{k},\bar{k}'\!+\bar{q}_P^0,\bar{k}',\bar{k}\!+\bar{q}_P^0) \! \cr
& \hskip 2 truecm  + \epsilon_{P,l}g_2(\bar{k}'\!+\bar{q}_P^0,\bar{k}, \bar{k}'\!,\bar{k}+\bar{q}_P^0)]\big\rangle,\\
&\mathrm{and}\cr
& \partial_\ell  z_{\rm BOW}(\bar{k}+ \bar{q}_P^0) =   {1\over 2\pi} \int dk'_\perp I_P(k_\perp',\bar{q}^0_P)   z_{\rm BOW}(\bar{k}' + \bar{q}_P^0)\cr 
&\! \! \times \! \big\langle  [ \epsilon_{P} g_1(\bar{k}'\!+\bar{q}_P^0,\bar{k},\bar{k}',\bar{k}\!+\bar{q}_P^0)  + \epsilon_{P,l}g_2(\bar{k}'\!+\bar{q}_P^0,\bar{k}, \bar{k}'\!,\bar{k}+\bar{q}_P^0) \! \cr
& - \epsilon_{P} g_3(\bar{k}',\bar{k}\!+\bar{q}_P^0,\bar{k}'\!+\bar{q}_P^0,\bar{k}) - \epsilon_{P,l} g_3(\bar{k},\bar{k}'\!+\bar{q}_P^0,\bar{k}',\bar{k}\!+\bar{q}_P^0)]\big\rangle \cr
\label{RGBOW}
\end{align}
for the  static $\mu=$SDW and BOW  susceptibilities, respectively,   at $\bar{q}_P^0=(\pi,0)$. In the superconducting channel, we shall examine 
 \begin{align}
\label{RGSC} 
& \partial_\ell  z_\mu(-\bar{k}+\bar{q}_C^0) =   {1\over 2\pi} \int dk'_\perp I_C(k_\perp',\bar{q}^0_C) z_\mu(-\bar{k}'+\bar{q}_C^0)\cr 
&\! \! \times \! \Delta_\mu(k_\perp)\big\langle \epsilon_C [ g_1(-\bar{k}'\!+\bar{q}_C^0,\bar{k}', -\bar{k}+\bar{q}_C^0,\bar{k})   \cr
&      \hskip 2 truecm + g_2(-\bar{k}'+\bar{q}_C^0,\bar{k}',\bar{k},-\bar{k}+\bar{q}_C^0)]\big\rangle,
\end{align} 
for the static SC susceptibility  at $\bar{q}_C^0=0$, where $\Delta_\mu(k_\perp)$  is the form factor for the SC  order parameter. For SC-{\it d} and triplet-f wave (SC-{\it f}) correlations, we have $\Delta_{\rm SC-{\it d}}(k_\perp)  = \sqrt{2} \cos k_\perp$ and $\Delta_{\rm SC-{\it f}}(k_\perp)  = ({\rm sgn} \,k)\sqrt{2} \cos k_\perp$, whereas for conventional singlet pairing (SC-{\it s}), we have $\Delta_{\rm SC-{\it s}}(k_\perp)=1$.

Before embarking on  the solution of the above equations, it is instructive  at this stage to examine their  basic  features as a function of the different energy scales of the model.  At high temperature  where $T\gg \omega_D$ and the  phonons are classical, the contribution of the Ph-M   interaction to  all   open diagrams--ladder and vertex corrections--becomes strongly dampened  for all $\Lambda(\ell)$, as a  result of  retardation that  reduces  the summations  over intermediate frequency transfer in   such diagrams. In this temperature range, the Ph-M  part contribute more appreciably to the closed-loop diagram  of the Peierls channel which does not have an intermediate sum over transfer frequencies, and this is  on equal footing with the direct Coulomb  part in Eqs. (\ref{RGg1})-(\ref{RGg3}). On the other hand,  when entering the-low  temperature domain at $T< \omega_D$,   retardation effects are reduced  which progressively  strengthen the contribution of the electron-phonon interaction to   open diagrams. This increases  mixing or interference between all diagrams of the Peierls and Cooper scattering channels. 

For the range of parameters considered in the model, the temperature scale $T_\mu$ of   instabilities of the electron gas that are considered below all fall in the temperature range $T_\mu\ll t_\perp$. This is where the transverse electron motion  and the warping of the Fermi surface are coherent, making the electron gas effectively two-dimensional, albeit strongly anisotropic in this temperature domain. This is  known to  affect the interference in a particular way depending on the energy distance $\Lambda(\ell)$ from the Fermi surface in the RG flow. At high energy, when  $\Lambda(\ell)\gg t_\perp$, the flow essentially coincides with the 1D limit where the interference is maximum, although  subjected to the above conditions between $T$ and $\omega_D$. When  $\Lambda(\ell)\ll t_\perp$, the interference between the Peierls and Cooper channels  is affected  by the coherent warping of the Fermi surface and ultimately  nesting alterations at   $\Lambda(\ell)< t_\perp'$. Both  generate a momentum dependence of the coupling constants (\ref{RGg1})-(\ref{RGg3}) which    reflects  in the end  the nature    of the electron gas instability at $T_\mu$.

\section{Results}
\subsection{Instabilities   for weak phonon-mediated interaction }
\begin{figure}
 \includegraphics[width=8.0cm]{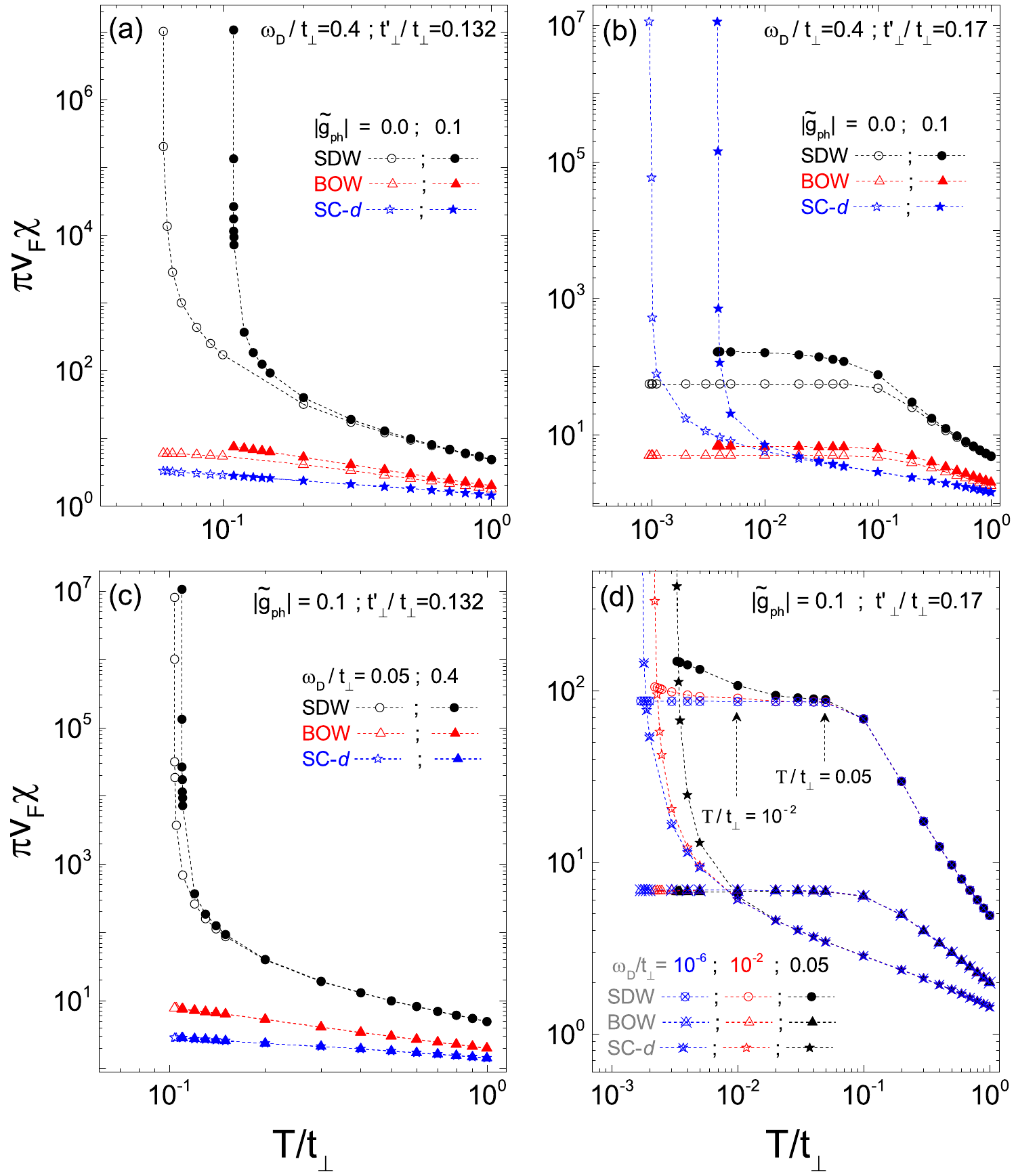}
 \caption{(Color on line) Typical temperature variations of the SDW, BOW and SC-{\it d} wave susceptibilities at $\omega_D/t_\perp=0.4$ for (a) weak    and   (b) intermediateantinesting  $t'_\perp$ at zero (open symbols ) and  a non-zero (solid symbols, $|\tilde{g}_{\rm ph}|= 0.1 $ )  phonon-mediated interaction. The comparison of susceptibilities for the same  $|\tilde{g}_{\rm ph}|$ for weak (c) and  intermediate (d) antinesting values at lower phonon frequencies.  \label{Responses1}} 
 \end{figure}

The integration of the  RG equations  up to $\ell \to \infty$ for   the couplings (\ref{RGg1})-(\ref{RGg3}) and   pair vertices (\ref{RGSDW})-(\ref{RGSC})  leads to the temperature dependence of the selected susceptibilities  as a function of antinesting, $t_\perp'/t_\perp$,  phonon frequency, $\omega_D/t_\perp$,  both normalized by the interchain hopping $t_\perp$; as for the  weak  Ph-M interaction, it is   parameterized by the ratio
\begin{equation}
\label{ }
|\tilde{g}_{\rm ph}| \equiv  | {g}_{\rm ph}|/g_1,
\end{equation}  
here normalized by the strength of non-retarded repulsive interaction $g_1$. The main features   the influence of weak   Ph-M coupling has on the  temperature dependence of relevant susceptibilities are summarized in  Fig.~\ref{Responses1}        at  small and intermediate antinesting parameter $t_\perp'$, and different $\omega_D$.  In Fig.~\ref{Responses1} (a),    $t_\perp'$  is taken sufficiently small so nesting   promotes  a singularity in $\chi_{\rm SDW}$, indicating an instability against the onset of  SDW order at $T_{\rm SDW}$. As for the correlations in the BOW and SC-{\it d} channels, the related susceptibilities are non singular and remain  small.   According to  Fig.~\ref{Responses1} (a), the presence of an even  small   $|\tilde{g}_{\rm ph}|$ at sizable $\omega_D$ is sufficient to cause a noticeable  increase
 of   $T_{\rm SDW}$  compared  to the purely electronic limit.

At the outset, the strengthening of SDW instability takes  its origin    in the $k$ and $q$ momentum dependence of Ph-M   interaction  in Eqs.~(\ref{g1})-(\ref{g3}) at $\ell=0$,   resulting in a reduction  of the backscattering and an increase of   the Umklapp term.  As discussed in more detailed in   Sec.~\ref{Isotope},   when Ph-M terms are small compared to  unretarded interactions, both concur to an increase of antiferromagnetic spin exchange between itinerant     electrons of opposite spins and located on separated sheets of the Fermi surface near $\pm k_F$. The above effects on scattering amplitudes are magnified by the RG flow,   mainly due to nesting in  one-loop  ladder and vertex corrections. Moreover,   the reinforcement of     SDW becomes the most efficient  in the temperature range $T<  \omega_D$ due  to the reduction  of retardation.  This is    where the Ph-M part acts  progressively as non retarded contributions in  all open diagrams such as the ladder and vertex corrections  of (\ref{RGg1}-\ref{RGg3}) and (\ref{RGSDW}),  namely those mainly involved in the exchange mechanism.    The  influence of Ph-M coupling on SDW correlations will  then naturally depend  on the value of  phonon  frequency $\omega_D$.   Figure~\ref{Responses1} (c)     shows indeed that  lowering  $\omega_D$ reduces the enhancement of  $T_{\rm SDW}$ at low antinesting,  an   indication of a positive isotope effect on SDW (see Sec.~\ref{Isotope}).

At   large-enough $t_\perp'$,       nesting  turns out to be sufficiently poor to prevent   the occurrence of    SDW.   The instability of the metallic state no longer takes place in the density-wave channel, but rather shows up by interference  in the Cooper channel with the onset of    SC-{\it d} order at $T_c$. As shown  in Fig.~\ref{Responses1} (b), the presence of a   small  Ph-M  coupling at the same     $\omega_D$ gives rise to a substantial increase of the critical temperature $T_c$ compared to the purely electronic case. The SC-{\it d} strengthening  is the mere consequence of  the boost of SDW spin fluctuations  responsible for the Cooper pairing in the metallic state. This is  shown  in Fig.~\ref{Responses1} (b)  where at non zero $|\tilde{g}_{\rm ph}|$ a more   pronounced, though non singular,  enhancement  of   $\chi_{\rm SDW}$ is found above $T_c$. This feature signals that the reinforcement of  spin fluctuations persists relatively deep in the normal state.

 In Fig.~\ref{Responses1} (d)  the effect of $\omega_D$ on both $T_c$ and normal state spin fluctuations is singled  out.  The growth of $T_c$  with $\omega_D$ is correlated with the increase of spin correlations above $T_c$.  In this part of the  Figure, we note that the onset   of spin fluctuation  reinforcement   takes place  at $T<\omega_D$ where  $\chi_{\rm SDW}$ clearly separates from  the static $\omega_D \to 0$ limit; it signals       the growth of ladder and vertex corrections following  a reduction of retardation. The enhancement of spin fluctuations in the normal phase will be analyzed in Sec.~\ref{CurieWeiss}, where it is found to  follow a  Curie-Weiss  temperature dependence, which is comparatively more pronounced than the one occurring in  the purely electronic limit \cite{Bourbon09,Sedeki12}. 
  
  Concerning BOW correlations, Fig.~\ref{Responses1} shows that for weak  $|\tilde{g}_{\rm ph}|$, these  remain weakly enhanced. However, as  will be shown next, the situation qualitatively changes  when  $|\tilde{g}_{\rm ph}|$, though still small,   reaches some critical value.  
 \begin{figure}
 \includegraphics[width=6.0cm]{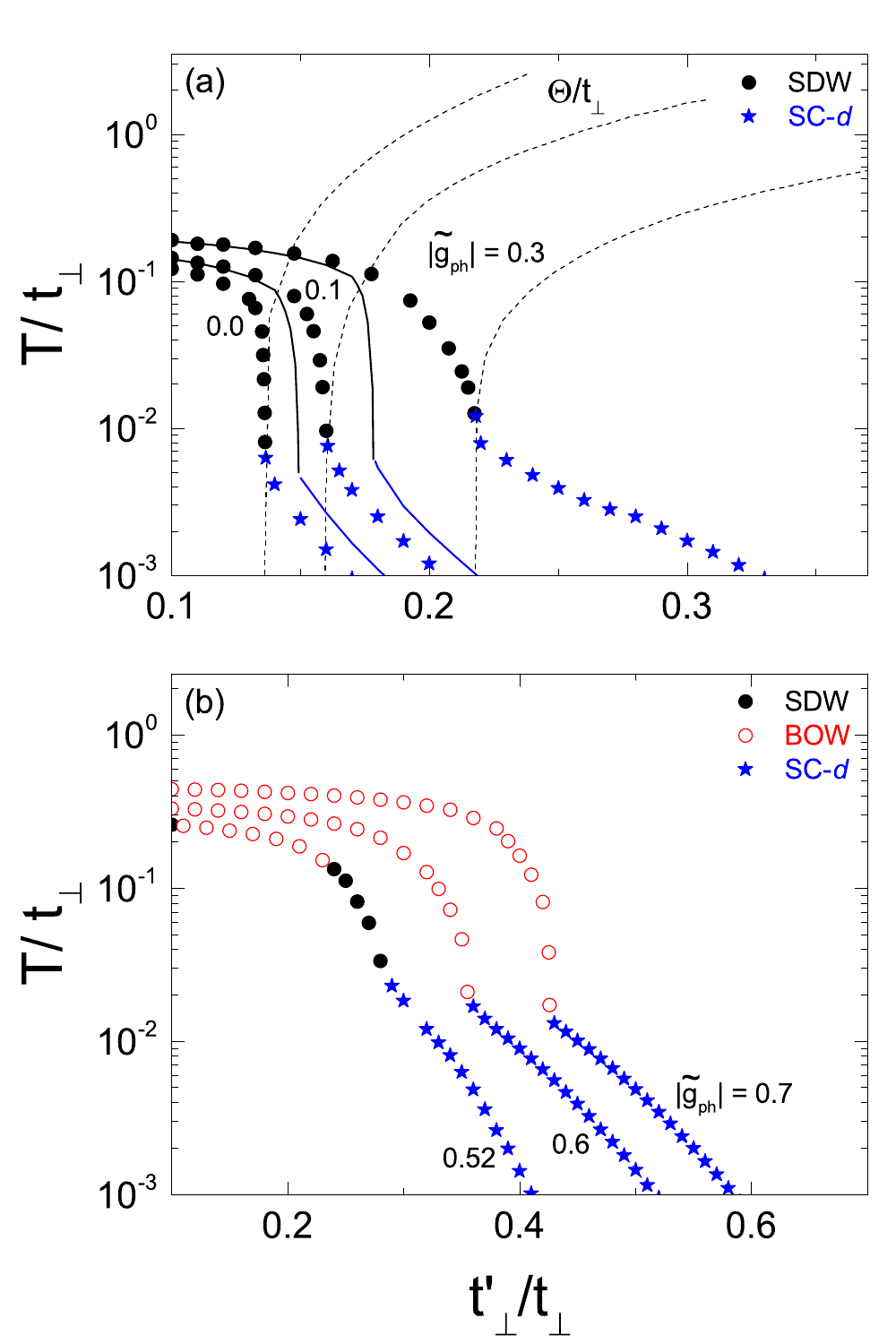}
 \caption{ (Color on line) Phase diagrams of the repulsive quasi-1D electron gas model as a function of  the anti nesting parameter $t_\perp'$  and   $|\tilde{g}_{\rm ph}| $ for (a), the SDW-SC-{\it d}    and (b)  BOW-SC{\it d} sequences of instabilities at $\omega_D/t_\perp=0.4$. In (a), the continuous lines correspond to the instability lines in the adiabatic $\omega_D\to0$ limit and the dashed lines show the variation of  the  Curie-Weiss  scale $\Theta$ of $\chi_{\rm SDW}$ [Eq.~(\ref{CW})]  as a function of $t_\perp'$ in the superconducting region.\label{Phases}} 
 \end{figure}
\subsection{Phase diagrams}

\subsubsection{Spin-density-wave  versus {\it d}-wave superconductivity}
We  now  consider   the  sequence of instabilities of the metallic state as a function of  $t_\perp'$ in order to   construct the phase diagrams at  weak Ph-M  couplings. This is shown in Fig.~\ref{Phases} (a).  At small     $|\tilde{g}_{\rm ph}|$ and  for a sizeable $\omega_D$,  the system remains  unstable  to the formation of a SDW state  with a $T_{\rm SDW}$ that displays the characteristic monotonic decrease  with increasing $t_\perp'$ \cite{Bourbon09,Sedeki12,Nickel0506,Duprat01,Yamaji82,Hasegawa86,Montambaux88}. At the approach of a well defined  antinesting threshold  $t_\perp'^*$,  however, $T_{\rm SDW}$ undergoes a critical drop  that terminates at $t_\perp'^*$  where  SC-{\it d}  begins  at its peak  value denoted by $T_c^*$. Above,  $ T_c$ shows a continuous decrease with $t_\perp'$  that  correlates with the reduction of SDW fluctuations as the source of Cooper pairing.

 As  stressed above,  Fig.~\ref{Phases} (a) confirms  that  the Ph-M  coupling, albeit small, reinforces both $T_{\rm SDW}$ and $T_c$ for all $t_\perp'$, including  the   critical value $t_\perp'^*$ at which superconductivity emerges. We also  note from Fig.~\ref{Phases} (a) that this reinforcement   reduces the  sharpness of its critical drop at the approach of  $t_\perp'^*$, an effect that carries over in the superconducting sector where the reduction of  $T_c$ with $t_\perp'$ turns to be less rapid. 
 
Also shown  in the figure    are the instability lines in the static, $\omega_D\to0$  limit [continuous lines of Fig.~\ref{Phases} (a)]. Retardation effects are found to be very  important at the approach of  the critical value $t_\perp'^*|_{\omega_D\to0}$ and beyond, an indication of that   the isotope effect is clearly non uniform as a function of $t_\perp'$ (see Sec.~\ref{Isotope}). It is also worth noticing  from the Figure that in the presence of dominant non retarded repulsive interactions, the   influence of Ph-M terms on both SDW and SC-{\it d} instabilities remains finite  in  the static limit. This contrasts with the situation  when only Ph-M  interactions are present,  and where $T_c\to 0$   as $\omega_D\to0$   for {\it s}-wave SC \cite{Bakrim10}.

\subsubsection{ Bond-order-wave versus superconductivity}

\begin{figure}
 \includegraphics[width=8.0cm]{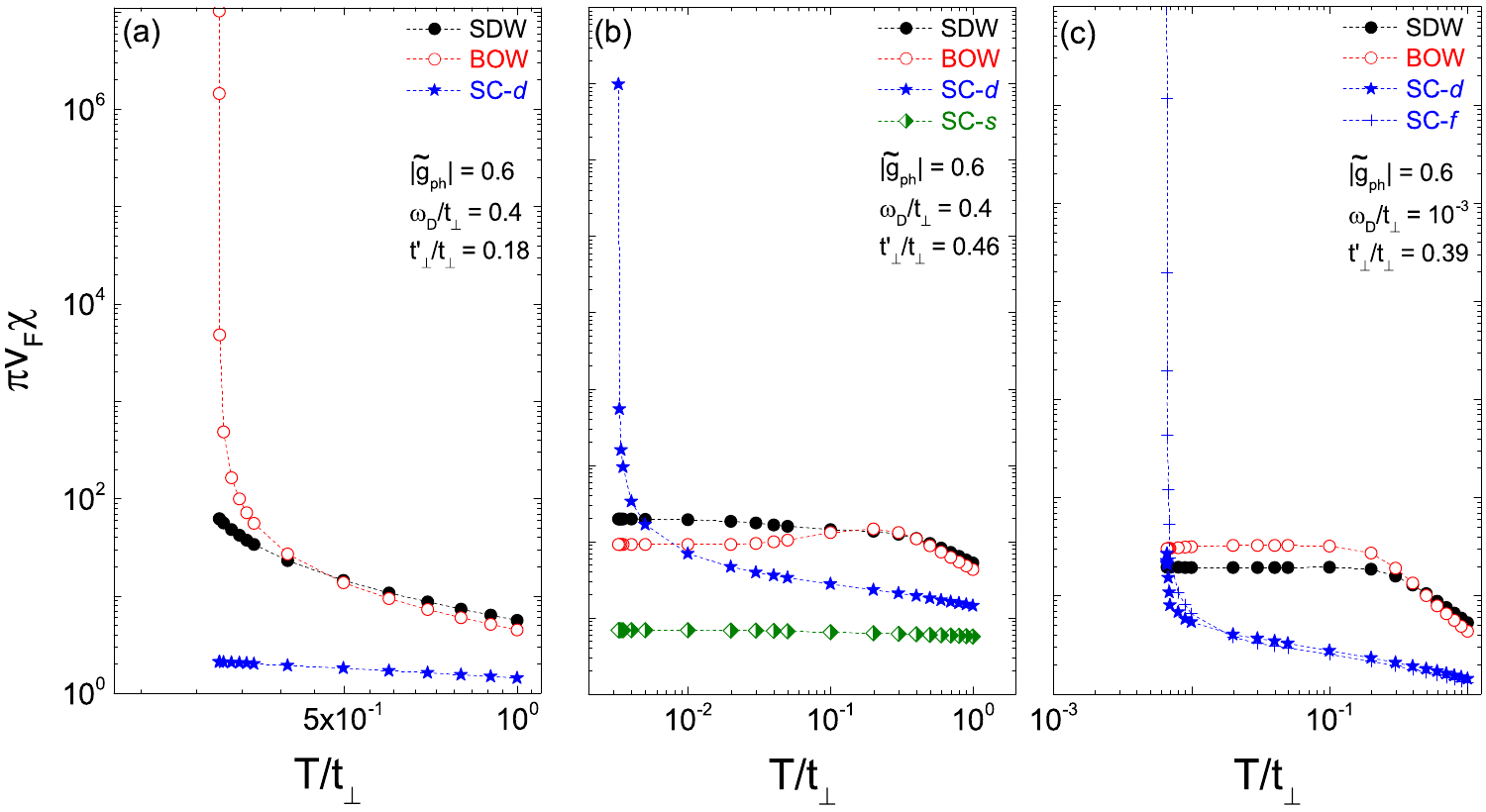}
 \caption{  (Color on line) Temperature variation of the SDW, BOW, and SC-{\it d} susceptibilities for  $|\tilde{g}_{\rm ph}|$ above the threshold $|\tilde{g}_{\rm ph}^{\rm c}|$ for  the occurrence of BOW instability at (a)  $t_\perp'<t_\perp'^*$, and in the superconducting sector at  $t_\perp'>t_\perp'^*$ for the (b)  SC-{\it d} ($\omega_D/t_\perp=0.4$) and (c) triplet SC-{\it f} ($\omega_D/t_\perp=10^{-3}$) instabilities. \label{Peierls}} 
 \end{figure}

\begin{figure}
 \includegraphics[width=8.0cm]{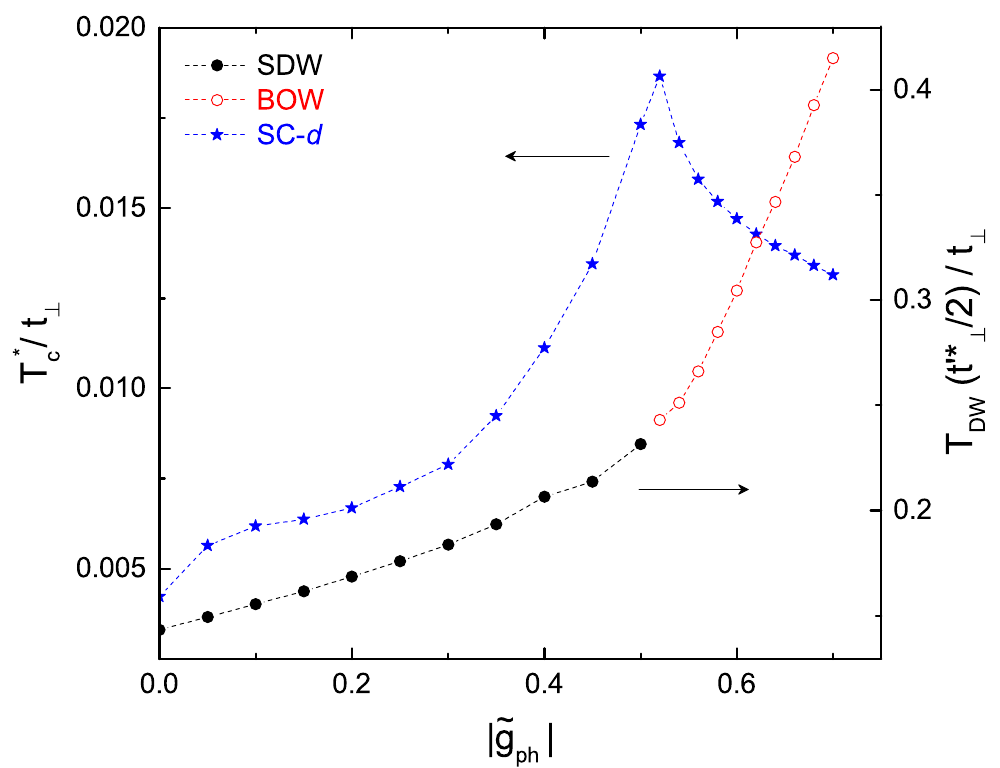}
 \caption{(Color on line) SDW/BOW critical temperatures at $t_\perp' =t_\perp'^*/2$ below the threshold antinesting (right) and the    maximum SC-{\it d} critical temperature (left) [$T_c^*=T_c(t_\perp'^*)]$   versus the normalized  strength of phonon-mediated  interaction  $|\tilde{g}_{\rm ph}|$ at $\omega_D/t_\perp=0.4$. \label{Tvsgph}} 
 \end{figure}

By increasing further  the strength of  Ph-M coupling for the same    $\omega_D$,  Fig.~\ref{Phases} (b) shows that the SDW-SC-{\it d} sequence of instabilities as a function of $t_\perp'$ is maintained only up to a  critical,    $|\tilde{g}^{\rm c}_{\rm ph}| (\approx 0.52$ for the parameters used),  above   which SDW turns out to be no longer stable and replaced by the onset  of a non-magnetic BOW state at $T_{\rm BOW}$. The  typical variations of relevant  susceptibilities in the  BOW sector of the phase diagram  are given in  Fig.~\ref{Peierls}-a. The BOW instability that takes place from the metallic state     corresponds to  the  onset of a   Peierls, though correlated,   lattice distorted state.  \cite{Caron84,Bakrim07}. 
%
 A   remarkable feature of the phase diagrams of  Fig.~\ref{Phases} (b)  is that   above $|\tilde{g}^{\rm c}_{\rm ph}|$ and at  $\omega_D$ that is not too small, the BOW instability  continues to be followed by SC-{\it d} superconductivity at $t_\perp'\ge t_\perp'^*$.   In  these conditions, however, $T_c$  becomes a decreasing function of $|\tilde{g}_{\rm ph}|$.  This is depicted  in Fig.~\ref{Tvsgph}, where it behaves so after  having reached its maximum at the  boundary   $|\tilde{g}_{\rm ph}^{\rm c}|$ where   SDW and BOW are found to be essentially degenerate and at their  maximum strength. It is worth noticing that at the boundary, $T_c$   has  increased by a factor 4 or so compared to the purely electronic case.  Despite the presence of a Peierls lattice distorted state,  the  essential role played by spin fluctuations in the emergence  of SC-{\it d} at $t_\perp'\ge t_\perp'^*$ remains. This is confirmed  in Fig.~\ref{Peierls} (b) where     $\chi_{\rm SDW}> \chi_{\rm BOW} $ over a large temperature interval at the approach of $T_c$ in the normal state.  In this sector we find no sign of increase for the {\it s}-wave superconducting correlations, as shown by the  temperature profile  of $\chi_{\rm SCs}$ that displays  no enhancement in Fig.~\ref{Peierls} (b). It is only  when $|\tilde{g}_{\rm ph}| \gtrsim 1$, that SC-{\it d} becomes in turn unstable and BOW ordering gives rise  to {\it s}-wave superconductivity under nesting alteration. The latter  case has been analyzed in detail by the same technique in Ref.~\cite{Bakrim10}.    

\begin{figure}
\includegraphics[width=6.0cm]{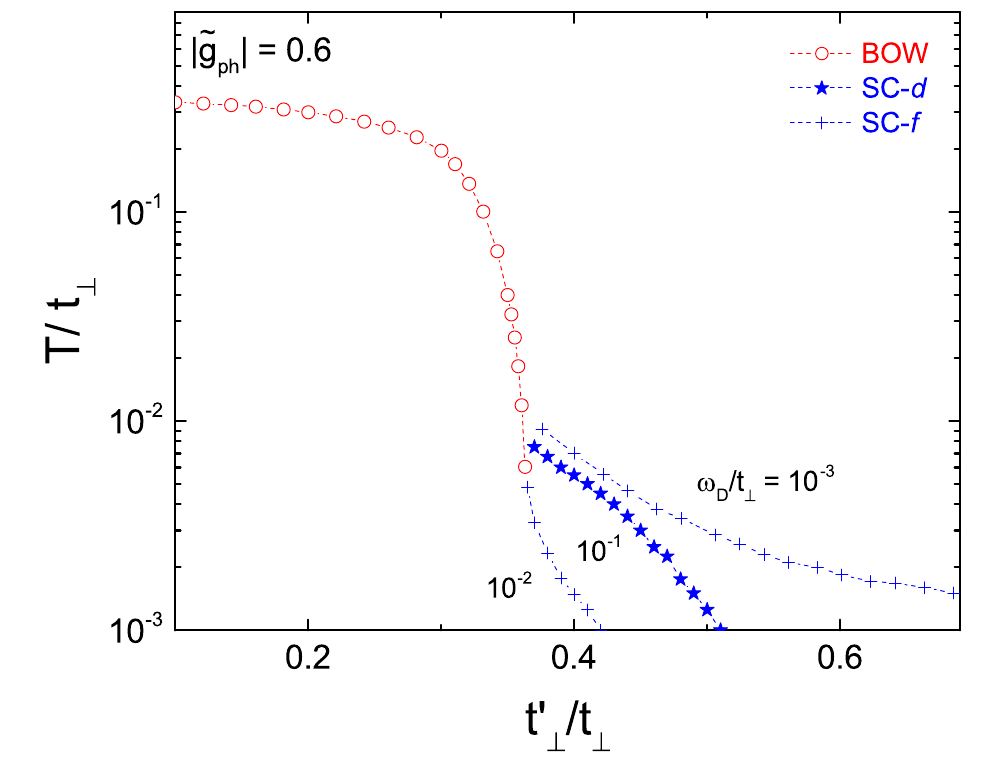}
 \caption{(Color on line) Phase diagram  above the threshold $|\tilde{g}_{\rm ph}^c |$ for the BOW-to-SC sequence of instabilities as a function of  antinesting.     The Figure shows the crossover between  triplet {\it f}-wave and  singlet {\it d}-wave  superconductivity in the small phonon frequency region. \label{Triplet} }
 \end{figure}

Another surprising feature of the phase diagram in  $ |\tilde{g}_{\rm ph}|>|\tilde{g}_{\rm ph}^{\rm c}|$ is found at low phonon frequency.  Figure~\ref{Triplet}   shows that in the small-$\omega_D$ range, the  BOW ordering at $t_\perp'\ge t_\perp'^*$ is followed by a triplet SC-{\it f} instability instead of a SC-{\it d} one. Since small phonon frequency increases retardation, it reinforces   most exclusively closed loop diagrams  in the RG flow,  related to density or charge fluctuations.  Bond charge correlations are then increased with respect to their spin counterpart; and for dominant repulsive interactions, this leads to SC-{\it f} type  superconductivity. The  triplet-singlet competition  is in a way  reminiscent of the  one found when  a weak repulsive (non retarded)  interchain interaction is added to the purely repulsive quasi-1D electron gas model\cite{Nickel0506}. The latter coupling is also  known to boost  exclusively charge fluctuations \cite{Lee77},   similar to the way the electron-phonon interaction does for closed loops  when strong retardation is present; the same interchain  coupling  is also known  to promote a SDW to BOW crossover in the density-wave instabilities at low antinesting\cite{Nickel0506}.
Cranking up $\omega_D$   results in the progressive enhancement of open diagrams which are  responsible for spin fluctuations and {\it d}-wave superconductivity. Although, from Fig.~\ref{Triplet}, the BOW ordering is weakly affected,  a SCf $ \to$ SCd crossover is indeed found to occur  at small $\omega_D/t_\perp $ ($\sim 0.1$ for the parameters used).

\subsection{Isotope effects}
\subsubsection{Spin-density-wave and {\it d}-wave superconductivity}
\label{Isotope}
In the preceding paragraphs we mentioned  on several occasions the positive  influence of raising     $\omega_D$ on the   strength of SDW and SC-{\it d} instabilities. This result  obtained  by varying  the molecular mass $M$ at fixed elastic  constant $\kappa$ [$g_{\rm ph}$  which remained constant according to  Eq.~(\ref{gph})],   corresponds to a positive  isotopic effect.   As touched on previously,  the mechanism of reinforcement of SDW correlation can be understood as a modification of the  effective antiferromagnetic exchange mechanism, itself affected    by retardation. Actually, for itinerant electrons, the total scattering amplitudes $g_2$ and $g_3$  of   the action $S_I$ in (\ref{SI}) contribute at $T$ an exchange term    of the form 
\begin{equation}
\label{Exchange}
S_{I}^{\rm ex} = \pi v_F {T\over L N_\perp}\sum_{\{\bar{\bf k}\},\bar{\bf q}_P} {1\over 2}(g_2 + g_3) \circ \vec{S}_{\bar{\bf k}_1,\bar{\bf q}_P}\cdot \vec{S}_{\bar{\bf k}_2,-\bar{\bf q}_P},
\end{equation}  
where  $\vec{S}_{\bar{\bf k},\bar{\bf q}_P}= {1\over 2} \psi^*_{+,\alpha}(\bar{\bf k}+ \bar{\bf q}_P) \vec{\sigma}^{\alpha\beta}\psi_{-,\beta}(\bar{\bf k}) +\, {\rm c.c} $ is the Fourier-Matsubara  component of the SDW spin density. Thus in weak coupling, the combination ${1\over 2}(g_2+g_3)$ corresponds  to  a momentum- and frequency-dependent antiferromagnetic  exchange interaction  generated by the scattering of oppositely moving carriers   at $\pm k_F$ with antiparallel spins. It is the same exchange term that  governs   the enhancement of the  vertex part $z_{\rm SDW}$ for the SDW susceptibility [See Eq.  (\ref{RGSDW})].  Its growth   with decreasing $\Lambda(\ell)$ results from the multiple exchange scattering of  virtual   ${\bf q}_P$ electron-hole pairs  carried by ladder and vertex corrections in the flow equations  (\ref{RGg2}) and (\ref{RGg3}).  As to   the  backscattering  term, $g_1$, its role  is indirect since   in the repulsive sector, $g_1$  tends to align spins of  $\pm k_F$  carriers.  This  dampens   the amplitude of both $g_2$ and $g_3$ and   then SDW correlations.
Therefore by lowering $\Lambda(\ell)$ the combined influence of a $g_1$ reduction and a $g_3$ increase by   Ph-M interactions in (\ref{g1}) and (\ref{g3}) will   boost $g_2$ and, in turn, $g_3$ and antiferromagnetic exchange.  
As mentioned earlier, however, this  additional and  positive  input  of Ph-M interaction  reaches its maximum impact in the temperature domain $T<\omega_D$, namely where retardation effects on virtual electron-hole pair scattering processes become small, hence the isotope effect on SDW.
 \begin{figure}
\includegraphics[width=8.0cm]{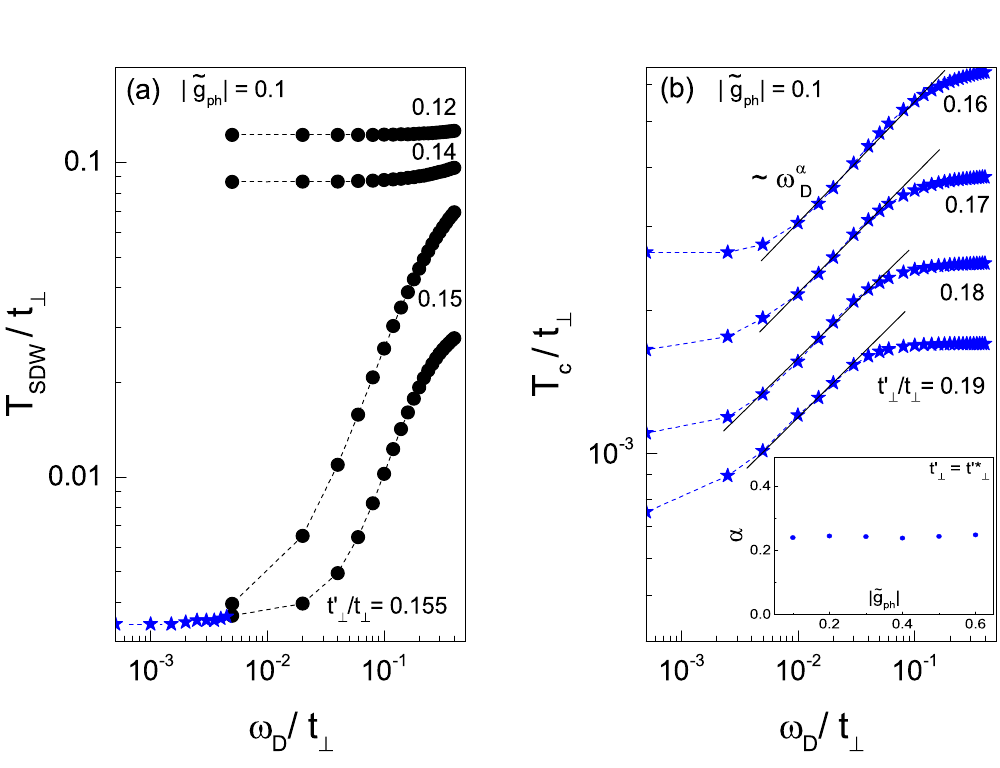}
 \caption{ (Color on line)  Isotope effect at  $|\tilde{g}_{\rm ph}|=0.1$ for (a) $T_{\rm SDW}$  at different anti nesting  $t_\perp' <  t'^*_\perp $   and   (b) $T_c$ of   the SC-{\it d} channel for different    $t_\perp'>  t'^*_\perp $. Insert: Variation of the isotope exponent as a function of phonon-mediated coupling amplitude at $t_\perp'^*$. }
 \label{FigIsotope}
 \end{figure}
The increase of $T_{\rm SDW}$ with $\omega_D$ is illustrated in Fig.~\ref{FigIsotope} (a) for   $|\tilde{g}_{\rm ph}|=0.1$  and different values of $t_\perp'$ in the SDW part of the phase diagram. At relatively small $t_\perp'$  that is, well into the SDW sector,    $T_{\rm SDW}$ undergoes a monotonic but weak increase over all the   frequency range of phonons, a consequence of ladder and vertex corrections to the antiferromagnetic exchange that grow in importance  by increasing  $ \omega_D$.    It is worth noticing that in  the adiabatic limit, $T_{\rm SDW}|_{\omega_D\to 0}$ is found to be  slightly larger than the $ T_{\rm SDW}|_{g_{\rm ph}=0} $ obtained in the absence of Ph-M interaction [see Fig.~\ref{Phases} (a)]. This indicates that static phonons still have a positive influence on the exchange interaction (\ref{Exchange}) and the strength of SDW correlations.
This adiabatic  effect finds a certain  echo in the strong coupling--Hubbard interaction--case where dynamical mean-field theory calculations   do predict an enhancement of antiferromagnetic exchange between localized spins by zero frequency phonons\cite{Sangiovanni05}. Here  the static enhancement essentially results from the mixing of Ph-M interaction to the  non-retarded Coulomb terms $g_i$ in the RG flow;  the enhancement vanishes  by taking  $g_i\to 0$ in  Eqs (\ref{g1}), (\ref{g2}), and (\ref{g3}), a result   found in the limit of pure   electron-phonon coupling \cite{Bakrim10}.

 When $t_\perp'$ increases and   approaches  the critical domain  where the   drop in $T_{\rm SDW}$ becomes   according to Fig.~\ref{Phases} (a), essentially  vertical,   the isotope effect   becomes huge as traced  in Fig.~\ref{FigIsotope} (a). Close to $t_\perp'^*$, the reinforcement of SDW correlations  by an even small increase in $\omega_D$  gives rise    a large  increase  of $ T_{\rm SDW}$. This is not the consequence of nesting improvement, but rather the result of stronger nesting deviations  needed to counteract the reinforcement of SDW instability   by Ph-M interactions.  For $t_\perp'$   slightly above 
$t_\perp'^*$,     Fig.~\ref{FigIsotope} (a)  features the interesting possibility  of  a SC-{\it d}-to-SDW transition    as a function of $\omega_D$. 

The positive isotope effect carries over into  the SC-{\it d} side of the phase diagram where   $T_c$ is  found to increase  with $\omega_D$  at different $t_\perp'$, as shown in Fig.~\ref{FigIsotope} (b). This is directly associated with the $\omega_D$- dependent reinforcement of spin correlations in the normal state as already pointed out in Fig.~\ref{Responses1} (d), which strengthens the pairing interaction in the SC-{\it d} channel. Although  the isotope effect is slightly larger in amplitude near  the critical $t_\perp'^*$, it remains of comparable size  at an arbitrary value of antinesting with a power law $T_c\sim \omega_D^\alpha$ that takes place   at an intermediate frequency with an exponent $\alpha \simeq 0.24 (\equiv d\ln T_c/ d\ln\omega_D)$, a value  virtually  independent of $t_\perp'$ [see Fig.~\ref{FigIsotope} (b)] and $|\tilde{g}_{\rm ph}|$, as shown in the insert of Fig.~\ref{FigIsotope} (b).
At high phonon  frequency where  the ratio $\omega_D/T_c$ becomes very large, retardation effects become negligible and $T_c$ tends to level off with frequency. This saturation probably reflects the limitation of using a finite number of Matsubara frequencies in the mean-field approximation of the loop  convolution over frequency. 
\subsubsection{Bond order wave  versus superconductivity}

\begin{figure}
\includegraphics[width=8.0cm]{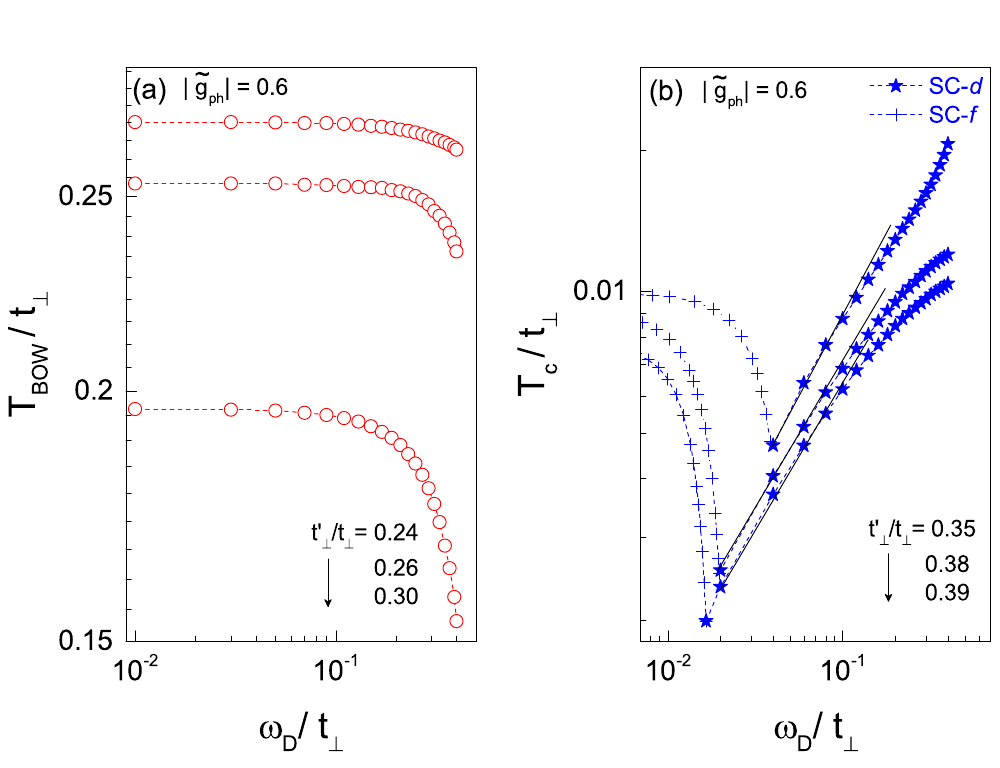}
 \caption{ (Color on line) Isotope effect at  $|\tilde{g}_{\rm ph}| > |\tilde{g}^c_{\rm ph}| $ for (a) $T_{\rm BOW}$  at different antinesting  $t_\perp' <  t'^*_\perp $   and on (b) $T_c$ in  the SC-{\it f} and SC-{\it d} channels for different    $t_\perp'>  t'^*_\perp $. The straight lines correspond to the power-law dependence $T_c \sim \omega_D^\alpha$, where $\alpha \simeq 0.25$. \label{IsotopeBOW} }
 \end{figure}

In the BOW regime above $|\tilde{g}^c_{\rm ph}|$, the isotope effect on  $T_{\rm BOW}$ has  the opposite sign. At low $t_\perp'$, for instance, Fig.~\ref{IsotopeBOW} (a) shows that $T_{\rm BOW}$ decreases monotonically with $\omega_D$ and the reduction becomes increasingly large with $t_\perp'$ which also softens  the lattice distortion through nesting alteration. A reduction of $T_{\rm BOW}$ with $\omega_D$ is a consequence of the growth of non adiabaticity of the phonon field, a well-known factor to be at play in the reduction of the Peierls distortion gap in purely electron-phonon models in one dimension\cite{Bakrim07,Caron84,Fradkin83}. From a diagrammatic point of view, non-adiabaticity is a quantum effect again tied   to the unlocking of  Ph-M interaction to open diagrams and  thus to quantum interference between electron-hole and Cooper pairing at the one-loop level. In contrast to the SC-{\it d}-SDW mixing, however, the interference is  in the present case destructive: Cooper  and Peierls  diagrammatic  contributions have opposite sign and this    reduces  the temperature scale of BOW ordering \cite{Caron84}.   The onset of  a quantum to classical crossover   for the BOW state  is perceptible at $\omega_D/2T^0_{\rm BOW}|_{\omega_D \to 0} \sim 1$, as it is found to occur in the pure electron-phonon limit\cite{Bakrim10,Bakrim07}. 

Above $t_\perp'^*$, but for small $\omega_D$, we still observe an inverse isotope effect for the $T_c$ of    triplet, SC-{\it f}   superconductivity, as shown in  Fig.~\ref{IsotopeBOW} (b). This confirms the role of BOW fluctuations in the existence of SCf ordering  at repulsive coupling. This is further supported when $\omega_D$ increases and crosses the critical value at which SC-{\it d} reappears in  Fig.~\ref{Triplet}. Then the isotope effect becomes once again positive as a consequence of the growth of antiferromagnetic exchange and spin fluctuations that govern the {\it d}-wave Cooper pairing. In the SC-{\it d} regime, one can extract at intermediate frequencies  a power law dependence $T_c\sim \omega_D^\alpha$ for the isotope effect with a value of $\alpha\simeq 0.25$ similar to the one found below $|\tilde{g}_{\rm ph}^c|$ [Fig.~\ref{FigIsotope} (b)].

\subsection{Normal state}
\label{CurieWeiss}
\begin{figure}
\includegraphics[width=8.0cm]{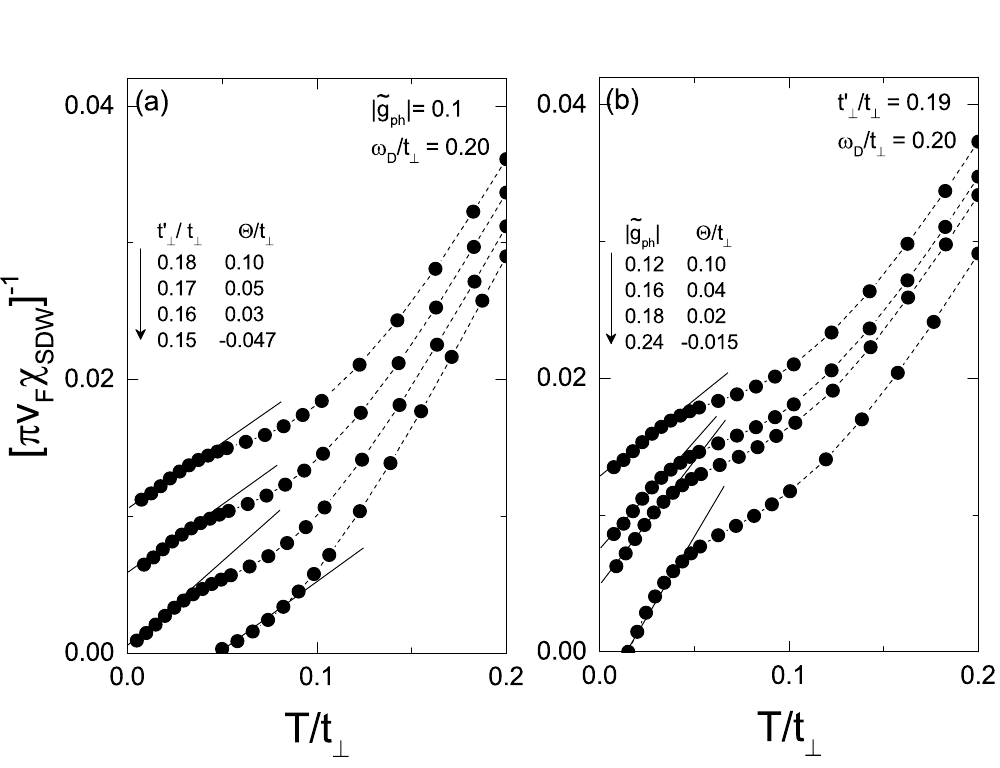}
 \caption{The temperature dependence of the normal phase inverse SDW susceptibility at  different antinesting  (a) and electron-phonon interaction strengths (b). The straight lines correspond to the Curie-Weiss fit [Eq.~(\ref{CW})]. \label{Normal} }
 \end{figure}

Now  that the positive influence of electron-phonon interactions on the temperature scales for ordering has been examined, one can
turn our attention on the influence of a weak Ph-M interaction on  spin correlations of the normal phase above $T_c$. This is done  for the SDW-SC-{\it d} sequence of instabilities.   In Fig.~\ref{Normal} (a), we show the temperature dependence of the inverse SDW susceptibility at small $|\tilde{g}_{\rm ph}|$ and various   strengths of antinesting. At sufficiently high $t_\perp'> t_\perp'^*$, $\chi^{-1}_{\rm SDW}$ decays essentially linearly from the high-temperature region and extrapolates  towards a critical point at a finite $T_{\rm SDW}$. However, as the temperature is lowered at $T< t_\perp'$, nesting deviations becomes coherent and  the susceptibility undergoes a change of regime and ceases to be critical.  Nevertheless, according to Fig.~\ref{Normal} (a),  $\chi^{-1}_{\rm SDW}$ keeps decreasing and extrapolates to a non zero intercept at $T=0$ and a finite slope at the end point  $T_c$. 

This non singular growth of spin correlations in the metallic state, which persist down to $T_c$, can be well described by a Curie-Weiss form (continuous lines in Fig.~\ref{Normal}):
\begin{equation}
\label{CW}
\chi_{\rm SDW} = {C\over T+ \Theta},
\end{equation}
extending up to the temperature $T_{\rm CW}$ for the onset of the Curie-Weiss regime, which is about 10 times $T_c$ in temperature at the frequency used in the figure \{$T_{\rm CW}$ decreases when $\omega_D$ is lowered [see Fig.~\ref{Responses1} (d)]\}. Here the Curie-Weiss scale $\Theta$ stands as a characteristic energy for SDW fluctuations, which is defined as positive when  $t_\perp'> t_\perp'^*$. The Curie-Weiss behavior has been already found in the purely electronic case   \cite{Bourbon09,Sedeki12}. It results from   the positive feedback of SC-{\it d} pairing on SDW correlations, a consequence of constructive  interference between these channels of correlations. The presence of Ph-M interactions clearly reinforces this behavior. As shown in Fig.~\ref{Normal} (b), cranking up $|\tilde{g}_{\rm ph}|$ leads to  the decrease of the Curie-Weiss scale $\Theta$, and an increase of the constant $C$ This  is consistent with an increase of the SDW correlation length $\xi\sim (T+ \Theta)^{-1/2}$, in tune with the increase of $T_c$ discussed above. The softening of $\Theta$ in Fig.~\ref{Normal} carries on until $t_\perp'$ reaches $t_\perp'^*$ where $\Theta \to 0$. There the system  would then become quantum critical with $\chi_{\rm SDW} \sim 1/T$  and  $T_{\rm SDW}\to0$, if not for the presence of superconductivity at a finite $T_c$ that prevents  the SDW quantum critical point from being reached. Below  $t_\perp'^*$, $\Theta <0$  and the system enters in the SDW sector with a finite $T_{\rm SDW} (\equiv -\Theta)> T_c$.

  At the approach of $t_\perp'^*$, $\Theta$ is well fitted by the quantum scaling form
\begin{equation}
\label{Theta}
\Theta \approx A (t_\perp'- t_\perp'^*)^\eta,
\end{equation}
with an exponent $\eta \simeq 1$, consistently with the product $\eta =\nu z$ of the correlation length ($\nu=1/2)$ and the dynamical ($z=2$) exponents for SDW at the one-loop level. The linear profile of $\Theta$ near $t_\perp'^*$ is illustrated  in Fig.~\ref{Phases}-a. From the  Fig.~2 (a) and Fig.~\ref{Normal} (b), the coefficient $A$ decreases relatively quickly with $|\tilde{g}_{\rm ph}|$. 

\section{Discussion and conclusion}

In this work we used a  weak-coupling RG  approach to examine the influence of the tight-binding electron-phonon interaction on the interplay between magnetism and superconductivity in quasi-one-dimensional correlated electron systems.  When the  phonon-mediated  interaction remains weak and subordinate to the direct Coulomb terms of the electron gas,    the RG flow of scattering amplitudes is  found to be distorted for particular  longitudinal electron momentum and momentum transfers. This    reinforces the  antiferromagnetic exchange mechanism between itinerant spins and yields   an increase of the temperature scale of  SDW ordering.  By introducing enough  nesting deviations into  the electron kinetics,  SDW ordering is inhibited, but magnetic reinforcement by the  electron-phonon interaction persists  and shifts by interference  in the superconducting channel. {\it d}-Wave Cooper pairing and   $T_c$ then become  enhanced compared to the purely electronic situation. These properties were found to be affected by   retardation effects linked  to the exchange of low energy acoustic phonons  that modulate  the strength  of  virtual electron-hole scattering entering   into the antiferromagnetic exchange term of the electron gas.  This gives rise to a positive isotope effect  on the SDW ordering temperature,  which carries over beyond  the critical  antinesting $t_\perp'^*$   where   {\it d}-wave superconductivity is found. 

Our results   also revealed that such an increase for   $T_c$   is preceded  by the  strengthening  of spin fluctuations in the normal phase.  This is manifest in a more pronounced   Curie-Weiss   SDW susceptibility compared to the purely electronic situation,  a  consequence   of self-consistency  between {\it d}-wave Cooper pairing  and spin fluctuations, a  positive interference effect whose amplitude scales with $T_c$.

We have also established the  range of  electron-phonon interaction beyond which  SDW ordering is no longer stable against the BOW or Peierls distorted state.   In these conditions, the Peierls ordering was  found to be followed above critical  antinesting  by  either {\it d}-wave or amazingly triplet {\it f}-wave superconductivity depending if retardation effects are weak or strong, respectively.  The isotope effect which is   negative in the triplet SC-{\it f} sector and positive in SC-{\it d} reflects the origin of the pairing interaction in both situations, namely BOW fluctuations in the former case and SDW ones in the latter.

The relevance of the above results for concrete materials showing the emergence of superconductivity on the verge of anti-ferromagnetism is of interest. In   Bechgaard salts  for instance,
 superconductivity  manifests itself where  SDW state ends under pressure. Their normal state is characterized  by important spin fluctuations over a large temperature interval above $T_c$  whose amplitude scales with the one of spin correlations under pressure,  as made abundantly clear  by NMR experiments\cite{Creuzet87b,Brown08,Wu05,Kimura11}. 
  
  Our findings show that intrachain  repulsive interactions are  dominant in these materials.  While repulsive interactions are known  to be   able to generate on their own the sequence  of SDW-SC-{\it d}  instabilities as a function  $t_\perp'$ in the quasi-1D electron gas model\cite{Duprat01,Nickel0506,Bourbon09,Sedeki12}, the present results show, however, that the addition of a relatively small  tight-binding electron-phonon interaction, which would  be compatible with diffuse x-ray scattering experiments\cite{Pouget12,Pouget82}, are  far from being an obstacle for superconductivity. When subordinate to the purely electronic repulsion,  the phonon-mediated  interaction  can  indeed play a very  active part  in assisting anti ferromagnetism in the emergence  of {\it d}-wave superconductivity with a stronger $T_c$. 
 
 Although the typical range of values taken by the electron-phonon matrix element has not been determined with great accuracy in materials like the Bechgaard salts (see, for example, Ref.~\cite{Pedron94}), the   results of the present paper suggest  that it should be small in amplitude compared to direct interactions. This is supported by the stability of the SDW state against the Peierls distortion, which, from the above results,  is found to be  assured only  within  a finite interval of weak phonon-mediated interaction at  essentially arbitrary retardation. Therefore the absence of the Peierls phenomena in the Bechgaard salts may be viewed as a mere consequence of the weakness of the electron-phonon coupling constant in these materials. This view would be consistent with previous estimations made from optics \cite{Pedron94} and also from the  fact that the only few materials showing   a lattice distorted phase   belong  to  the more correlated   isostructural compounds of the (TMTTF)$_2X$ series, the so-called  Fabre salts.  A compound like (TMTTF)$_2$PF$_6$, for instance, is well known to undergo a spin-Peierls transition within a strongly correlated Mott state\cite{Pouget82,Pouget12,Creuzet87}. Less than 10 kbars of pressure is  sufficient  to weaken the coupling of phonons to  electrons  and transform this state into  one with antiferromagnetic N\'eel order \cite{Caron88,Chow98};  30 kbars separate  the latter    from  the sequence of SDW-SC instabilities found in  the prototype compound (TMTSF)$_2$PF$_6$  of the Bechgaard salts \cite{Wilhelm01,Adachi00,Moser98},  in line with a coupling to   phonons   that remains in the background of   direct Coulomb terms.
  
As to   the  possible experiments able to disentangle the part played by phonon-mediated  interaction on the SDW-SC sequence of instabilities seen in molecular materials like the Bechgaard salts,  isotope effect measurements would be certainly  of interest, especially near the quantum critical point   where the present results show that it    becomes huge at the approach of $t_\perp'^*$ on the SDW side of the phase diagram.  While the isotope effect in molecular materials proves to be difficult to realize in practice  due to the complications of controlling  all other parameters   following a change in  the mass $M$ of molecular units (volume of the unit cell, disorder,  etc.), the $^{13}$C enrichment of  the  TMTSF molecular units stands probably as the best way to limit these side effects and  to test some of the results obtained here.  According  to Fig.~\ref{Phases} (a), for instance,   a finite reduction in $\omega_D$  would  induce  a decrease in the critical $t_\perp'^*$ at which superconductivity occurs. Practically, one should therefore  expect  a  downward shift of the critical pressure for the emergence of superconductivity and a decrease in the maximum $T_c^*$  at that point and beyond on the pressure axis.
 
 Another possible signature of the reinforcement of anti-ferromagnetism by  electron-phonon interaction  in the Bechgaard salts  may  be found in  its  influence on the  Curie-Weiss behavior of SDW susceptibility which governs the   enhancement of the NMR spin-lattice relaxation rate observed  down to $T_c$  \cite{Brown08,Creuzet87b,Wzietek93,Wu05,Shinagawa07}. While the quasi-1D  electron gas model with purely electronic interactions   does predict a critical linear suppression of the Curie-Weiss  scale $\Theta$  for spin fluctuations as $t_\perp' \to t_\perp'^*$ \cite{Bourbon09,Sedeki12}, its slope [coefficient $A$ of   Eq.~(\ref{Theta})] proves to be  significantly  larger  than the one seen in   experiments \cite{Bourbon11}. In this regard, we have found that  adding a small $|\tilde{g}_{\rm ph}|$  is sufficient  to reduce the downslope of $\Theta$  to values congruent with experiments \cite{Bourbon11}, and this  over a large range  of retardation. This  supports  the view of an active role  played  by the electron-phonon interaction     in the properties of the metallic state, especially those associated to quantum criticality at $t_\perp'^*$.
 
 In this paper, we have dealt exclusively with the coupling of correlated electrons to low-energy acoustic  phonons within the tight-binding scheme for the electronic structure,  a coupling well known   to  be responsible for  electronically  driven  structural instabilities in low dimensional molecular materials \cite{Pouget12,Su80B}. We did not consider  intramolecular (Holstein) phonon modes,   also well known to be present.   
 Their classification alongside   their (small) coupling to electrons in (TMTSF)$_2X$ have been obtained from infrared optical studies \cite{Pedron94} .These molecular phonons are characterized by relatively large energies and weak retardation effects  compared to acoustic branches considered above. In first approximation, their influence can then be incorporated through a redefinition of the non retarded  terms,   amounting to a small and similar downward shift of the couplings $g_i$ of the electron gas model. Since the latter couplings was taken as phenomenological constants whose range  were fixed by experiments, the  values taken in the present  work should embody to some extent the influence of intramolecular phonons.
 
 The interplay between electron-phonon and electron-electron interactions in the framework of the Holstein-Hubbard model has been the subject of considerable attention in the past few years, especially in one dimension   where in the absence of interchain hopping and nesting alteration, SDW  order is found to  compete exclusively with a charge-density-wave state and to a lesser degree {\it s}-wave superconductivity when the phonon-mediated interaction strength is of the order of the direct Coulomb term \cite{Hardikar07,Tam07b,Bauer10b,Nowadnick12}.

 In conclusion, we have performed  a finite-temperature renormalization group analysis of the quasi-1D electron electron gas model with nonretarded electron-electron couplings and  phonon-mediated interactions of the  tight-binding electronic structure. For a phonon-mediated interaction that is   weak compared to  non retarded terms, we found a reinforcement of anti ferromagnetism and its transition toward  superconductivity   under bad nesting conditions of the electron spectrum.  The weakness of phonon-mediated interactions acts as a decisive factor for the stability of anti ferromagnetism against the Peierls phenomena in low-dimensional conductors. It is   likely that these retarded interactions  also have    a built-in positive impact in the observation of organic superconductivity on the verge of anti ferromagnetism in  the Bechgaard salts.

 \acknowledgments
 C. B. thanks the National Science and Engineering Research Council  of Canada (NSERC) and the R\'eseau Qu\'eb\'ecois des Mat\'eriaux de Pointe (RQMP) for financial support. Computational resources were provided
by the R\'eseau qu\'eb\'ecois de calcul de haute performance
(RQCHP) and Compute Canada.
  \bibliography{/Users/cbourbon/Dossiers/articles/Bibliographie/articlesII.bib}
\end{document}